\documentclass{aa}
\usepackage{psfig}
\begin{document}

\title{{\it XMM-Newton} observations of the BL Lac MS 0737+7441\thanks{Based on observations with XMM-Newton, an ESA Science Mission 
    with instruments and contributions directly funded by ESA Member
    States and the USA (NASA)}
    }

\author{Th. Boller, \inst{1}
M. Gliozzi, \inst{1} 
G. Griffiths, \inst{2} 
S. Sembay, \inst{2}
R. Keil, \inst{1}
O. Schwentker, \inst{1} 
W. Brinkmann, \inst{1} \and
S. Vercellone \inst{3}
}

\offprints{Th. Boller}

\institute{Max-Planck-Institut f\"ur extraterrestrische Physik,
Postfach 1603, 85748 Garching, Germany \\
\email{bol@mpe.mpg.de}
\and
X-ray Astronomy Group; Department of Physics and Astronomy;
Leicester University; Leicester LE1 7RH; U.K.
\and
Istituto di Fisica Cosmica "G. Occhialini", CNR, Milano, Italy 
}

\date{Received October 2, 2000; Accepted October 26, 2000 }

\abstract{
We report on the {\it XMM-Newton} observations of the 
BL Lac object MS~0737.9+7441 during the performance verification 
phase. A simple power--law fit provides an adequate description of the 
integrated spectrum in the 0.2--10 keV energy band. The photon index  is 
slightly steeper in the EPIC pn data with $\Gamma = (2.38 \pm 0.01)$ compared
to the EPIC MOS data ($\Gamma = (2.28 \pm 0.01)$). The difference is most
probably due to the present uncertainties in the calibration of the
EPIC MOS and EPIC pn data sets. We report evidence for intrinsic absorption
in the distant BL Lac above the Galactic column 
($N_{\rm H, Gal} = 3.2 \cdot 10^{20}
\rm{cm^{-2}}$) which is 
$N_{\rm H, fit}^{\rm z=0.315} \rm
= (2.70 \pm 0.20) \times 10^{20}\ cm^{-2}$ in the EPIC pn data and
$N_{\rm H, fit}^{\rm z=0.315} \rm
= (3.25 \pm 0.25) \times 10^{20}\ cm^{-2}$ in the EPIC MOS data assuming
neutral gas and solar abundances.
The flux  variations are found to be of the order of 10 \%.
No significant spectral variability is detected. 
}

\maketitle

\markboth{Th. Boller et al.:{\it XMM-Newton} observations of MS 0737+7441}{Th. Boller et al.:{\it XMM-Newton} observations of MS 0737+7441}

\section{Introduction}

The BL Lac object MS 0737.9+7441 was discovered in the {\it Einstein}
Observatory Extended Medium-Sensitivity Survey (EMSS; Gioia et
al. 1990, Stocke et al. 1991) with a flux of $\rm f_X = (9.8 \,\pm 0.6) \cdot
10^{-12} \ erg \ cm^{-2} \ s^{-1}$ in the energy range
between 0.3 and 3.5 keV.
Its redshift is z = 0.315 (Morris
et al. 1991).
In the 
ROSAT All-Sky-Survey observations the source had a count rate of 0.49
counts s$^{-1}$ during an exposure of 456 seconds.
Perlman et al. (1996) 
obtained a 
best fit power--law photon index of $\rm \Gamma = 1.91$ from the observed
hardness ratios assuming Galactic absorption $N_{\rm H, Gal} = 3.2 \cdot 10^{20}
\rm{cm^{-2}}$. Lamer et al. (1996) examined pointed ROSAT-PSPC data of MS
0737.9+7441 with a count rate of 0.51 $\pm$ 0.01 counts per second for
an exposure of 8782 seconds. Using a power--law model the
best fit parameters are  $N_{\rm H, fit} = (4.16
\pm 0.48) \cdot 10^{20} \rm{cm^{-2}}$ and $\Gamma = 2.39 \pm 0.11$.
In a {\it BeppoSAX} observation (Wolter et al. 1998) MS 0737.9+7441 was detected
in the LECS instrument with 37.1 $\pm$ 7.8 net counts during an exposure
of 3075 seconds, the MECS detector net  counts were 735.9 $\pm$ 30.6 
in 23279 seconds. By assuming a simple power law 
the 
best fit
parameters are $\Gamma = 2.53^{+0.28}_{-0.23}$ and 
$N_{\rm H, fit} = (25.8^{+49.3}_{-21.6}) \cdot 10^{20} \rm{cm^{-2}}$. 
A broken power
law resulted  in photon indices of $\Gamma{_1} = 1.17$ (which they classify
as uncertain) and
$\Gamma{_2} = 2.43^{+0.18}_{-0.16}$. The  break energy is 
1.05 (1.27--1.61) keV.

In the following we report on the {\it XMM-Newton} observation of 
MS~0737.9+7441 obtained during the performance verification program. 
Two exposures of approximately 20 ksec and 60 ksec were performed on the 
source between April 12, 2000 and April 13, 2000, either side of the orbital 
apogee gap. Around the middle of the second PN exposure the camera suffered a 
short telemetry break. The PN camera was in full frame mode throughout the
observation. Both MOS cameras were in large window ($300 \times 300$ pixels)
mode. All cameras employed their respective thin--1 filters.

The Hubble parameter was chosen to be $H_0 = \rm 70\ km\ s^{-1}\ Mpc^{-1}$
and a cosmological deceleration parameter of $q_0 = \rm \frac{1}{2}$ have
been adopted throughout the paper.

\section{Spectral fitting results}

\subsection{EPIC pn results}

\begin{figure}
%\centerline{\psfig{figure=spec_merged.ps,width=8.0cm,angle=-90,clip=}}
\centerline{\psfig{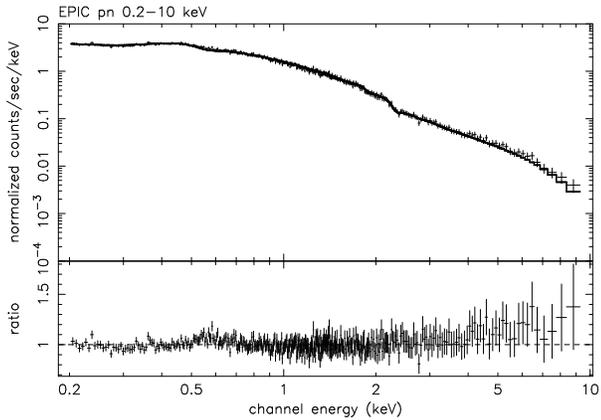}}
      \caption{Power-law fit to the EPIC-pn data. 
The upper panel shows the count rate distribution of MS~0737.9+7441 
(marked by the crosses) in the EPIC pn detector versus energy. 
The solid line gives the model spectrum. The lower panel gives the 
ratio between the data and the  model.}
\end{figure}

The spectral fitting results described below were obtained from the merged 
data set. Only single events were used from an extraction radius
of 75 arcsec  around the source position, the background was taken from the 
same chip with an extraction radius of 110 arcsec. 
A simple power law fit with the Galactic column fixed to
$N_{\rm H, Gal} = 3.2 \cdot 10^{20}
\rm{cm^{-2}}$ and allowing the intrinsic absorption and the photon 
index to be free parameters 
provides an adequate fit to the {\it XMM-Newton} data 
($\chi^2$ = 375 for 380 d.o.f.; cf. Figure 1).
We have used the {\tt XSPEC} models {\tt zphabs} to access the intrinsic
absorption in the source and {\tt phabs} for the Galactic absorbing 
column density.
The intrinsic absorbing column density measured with the EPIC pn  is 
$N_{\rm H, fit}^{\rm z=0.315} \rm
= (2.70 \pm 0.2) \times 10^{20}\ cm^{-2}$ assuming neutral gas
and solar abundances, consistent with the excess absorption suggested by the 
ROSAT data.
The photon index is $\Gamma$=2.38 $\pm$ 0.01.
The errors correspond to 90\% confidence levels for 1 
interesting parameter.
Using the F-test for the addition of one free parameter 
one gets $\rm \Delta \chi^2/\chi_{\nu}^2$ = 146 (cf. equ. 11.50 of
Bevington \& Robinson 1992). According to Table C.5 of Bevington \& Robinson 
(1992) this corresponds to a highly significant improvement 
($>$99.99 per cent)
of the fit quality
with intrinsic absorption  
compared to the fit with no intrinsic absorption. 
The mean 0.2--10 keV absorbed flux obtained from the {\it XMM-Newton} 
observations 
is $f \rm = 8.6 \times 10^{-12}\ erg\ cm^{-2}\ s^{-1}$. The
unabsorbed flux is $f \rm = 1.3 \times 10^{-11}\ erg\ cm^{-2}\ s^{-1}$,
corresponding to an isotropic luminosity 
of $L_X \rm = 3.6 \times 10^{45}\ erg\ s^{-1}$.
Spectral residua appear between about 0.5 and 0.7 keV in the EPIC pn
spectrum (cf. Figure 1). These wiggles might be attributed to
uncertainties in the presently available EPIC pn response matrix not allowing us to make definitive statements 
on the possible presence of soft X-ray emission lines, 
e.g. emission due to O~VIII (653 eV) and Fe~XXVII (726 eV) would 
fall in the energy range of the spectral residua. Better 
calibration and/or deeper observations are needed to further 
settle  this issue.
We 
note that these features do not significantly influence the spectral continuum
shape. 

\subsection{EPIC MOS results}

In the following the spectral fitting results to the combined MOS~1 and
MOS~2 observations are presented (cf. Figure 2). All events within the MOS X--ray pattern
library were used ({\it i.e.} Patterns 0 to 12). As with the pn, the MOS data 
in the 0.2 to 10 keV band are well--fit by a single power law with neutral 
absorption somewhat higher than the nominal Galactic value. A power--law 
model with the Galactic absorbing column fixed and adding a redshift corrected 
neutral hydrogen absorbing column density component gives, 
$N_{\rm H, fit}^{\rm z=0.315} \rm = (3.25 \pm 0.25) \times 10^{20}\ cm^{-2}$ 
and $\Gamma$ = 2.28 $\pm$ 0.01. The errors are 90\% confidence for 1 
interesting parameter. The reduced $\chi^2$ value is 1.27 for 
531 d.o.f. No significant systematic difference was found fitting the
data from the MOS cameras separately. The residuals to the MOS fit reveal 
correlated variations which are at most discrepant by 10\% and in general
near to the major instrumental absorption edges at Carbon, Oxygen and 
Silicon. The measured flux is consistent within 5 \% of the pn value.

\begin{figure}
%%\centerline{\psfig{figure=spec_mos.ps,width=8.0cm,angle=-90,clip=}}
\centerline{\psfig{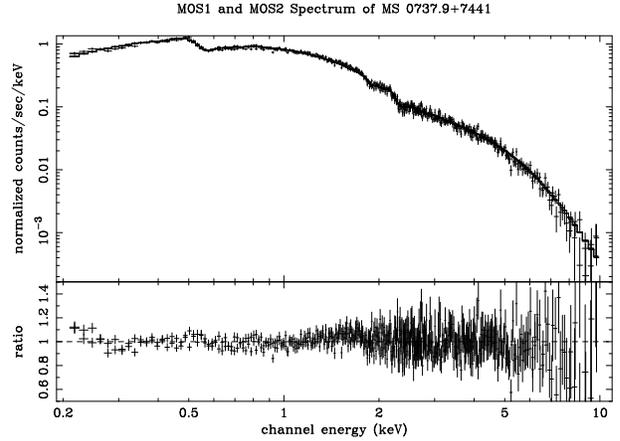}}
      \caption{Power-law fit to the EPIC MOS 1 and MOS 2 data.
The upper panel shows the count rate distribution of 
MS~0737.9+7441 (marked by the crosses) in the EPIC MOS 1 and MOS 2 detectors 
versus energy. The solid line gives the model spectrum. The lower panel 
gives the ratio between  the data and the model.}
\end{figure}

\begin{figure}
%\centerline{\psfig{figure=shB-1000.ps,width=8.0cm,angle=90,clip=}}
\centerline{\psfig{figure=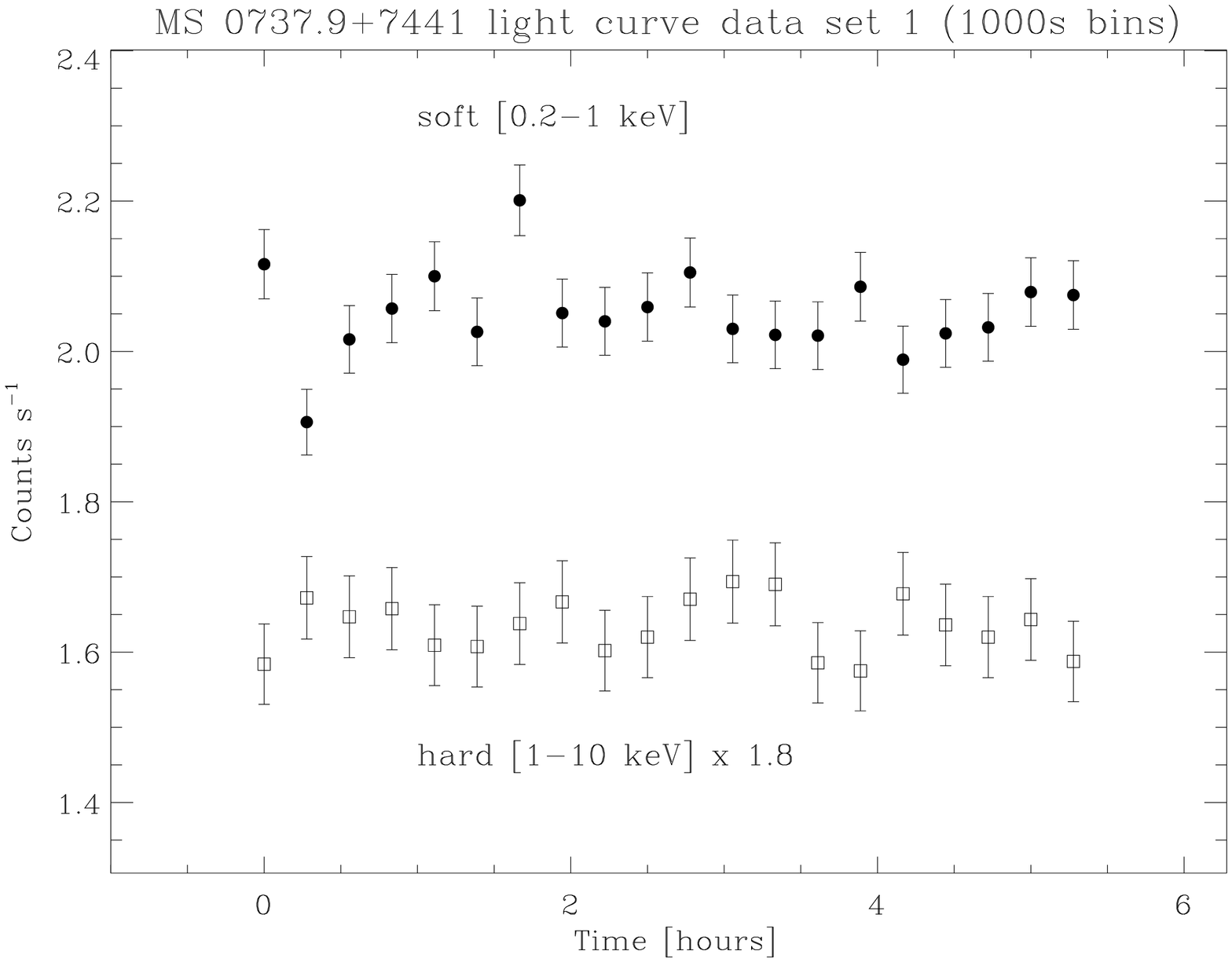,width=8.0cm,angle=90,clip=}}
%\centerline{\psfig{figure=shA-1000.ps,width=8.0cm,angle=90,clip=}}
\centerline{\psfig{figure=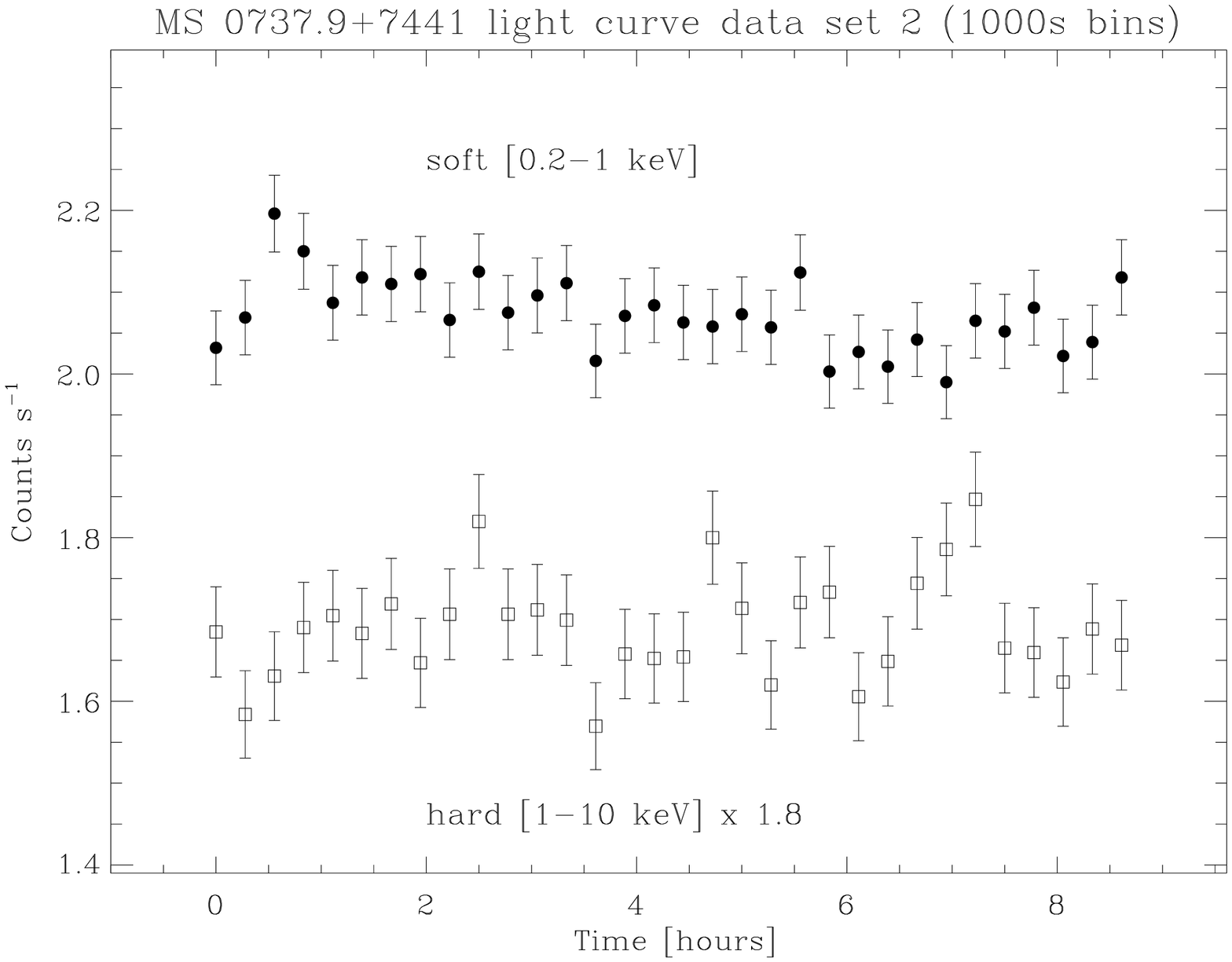,width=8.0cm,angle=90,clip=}}
%\centerline{\psfig{figure=shC-1000.ps,width=8.0cm,angle=90,clip=}}
\centerline{\psfig{figure=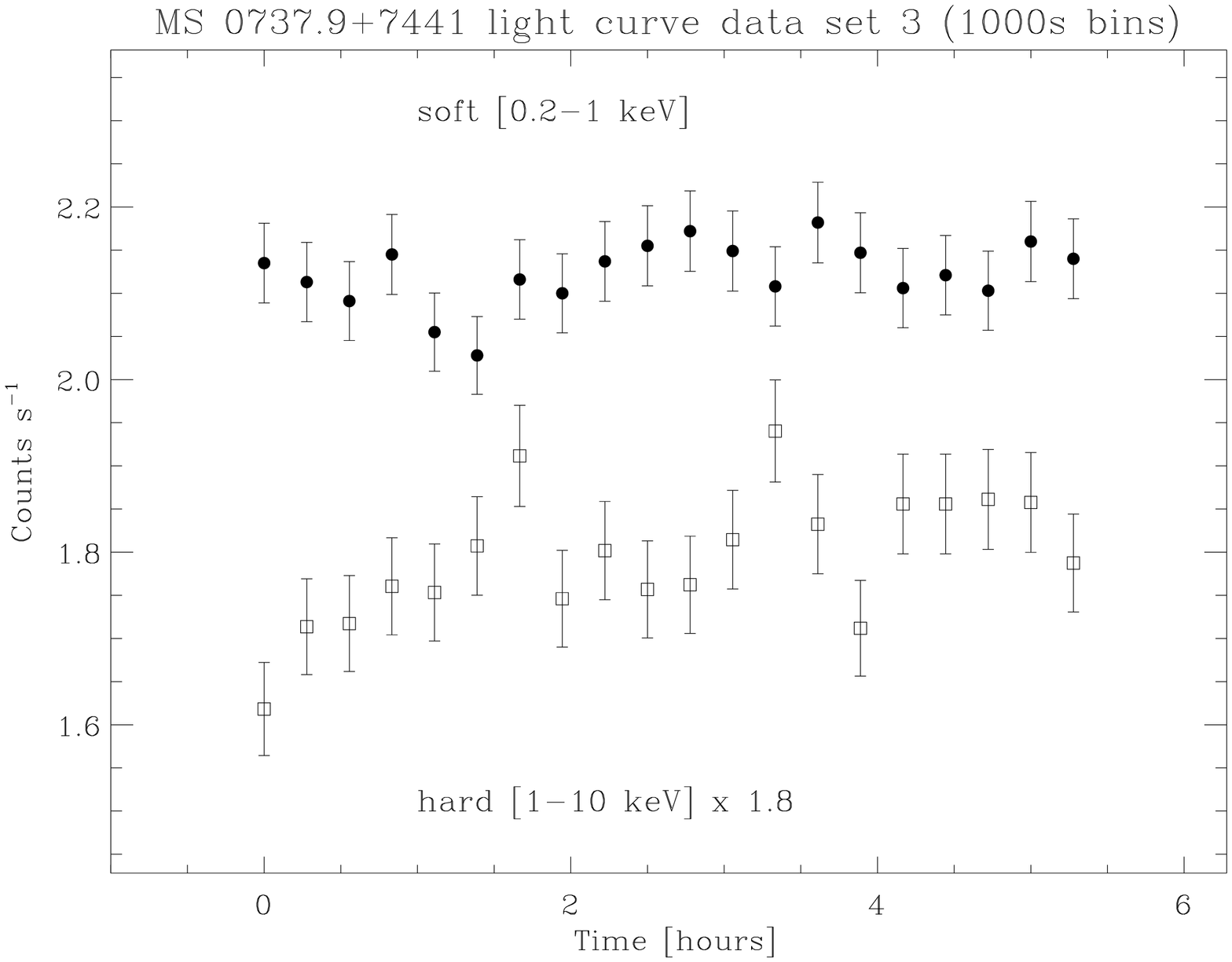,width=8.0cm,angle=90,clip=}}
      \caption{EPIC pn light curve for the three continuous observation
segments in different energy bands.}
\end{figure}

\begin{figure}
%\centerline{\psfig{figure=ratio.ps,width=8.0cm,clip=}}
\centerline{\psfig{figure=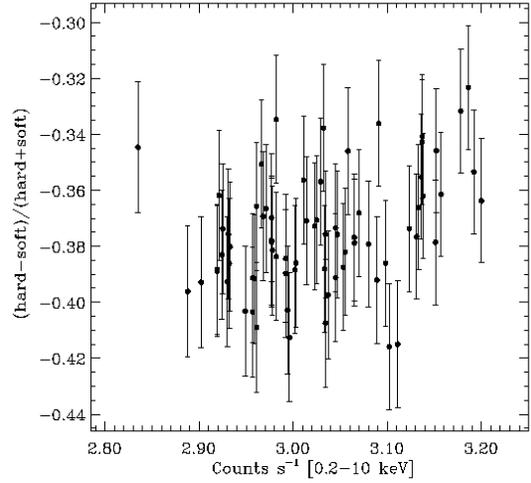,width=8.0cm,clip=}}
      \caption{Hardness ratio obtained from the soft (0.2--1.0 keV) and the
hard (1.0--10 keV) energy band ratio versus count rate. No significant spectral
variability is detected.}
\end{figure}

\section{Timing properties}

In Figure 3 we show the soft (0.1--1 keV) and hard (1--10 keV) light curves
for the three continuous observation segments on MS 0737.9+7441. 
The amplitude variability does not exceed a factor of about 10 \%, both in 
the soft and the hard band.
None of the X-ray variability events shown in Figure 3 exceed the radiative
efficiency limit (Fabian 1979; Brandt et~al. 1999), sometimes observed in 
BL Lac  objects.

\section{Search for spectral variability}

While X-ray amplitude variations of about 10 \%  are present in the EPIC 
light curves, these flux variations are not correlated with significant
spectral variability. No significant difference was found in the variability
behavior as seen by the pn and MOS cameras, so for brevity we present the
pn results only. In Figure 4 we plot the hardness ratio obtained from 
the soft (0.2--1 keV) and hard (1--10 keV) energy band in the pn camera 
versus the count rate. The hardness ratio remains constant within the errors.

\section{Summary}

XMM-Newton observations reveal that a featureless 
simple power law model
with absorption by neutral hydrogen 
provides an adequate fit to the
data. The photon indices as measured with the EPIC pn and EPIC MOS
are slightly different with 2.38 $\pm$ 0.01 and 2.28 $\pm$ 0.01,
respectively, most probably due to present calibration uncertainties for
the EPIC detectors.
The spectral residua between 0.5 and 0.7 keV might be attributed to uncertainties in the presently available EPIC pn response matrix not allowing us to make definitive statements on the possible presence of soft X-ray emission lines, e.g. emission due to O~VIII (653 eV) and Fe~XXVII (726 eV) would 
fall in this energy range.
We confirm the presence of intrinsic absorption in the source,
which is about 
$N_{\rm H, fit} = 3.0 \cdot 10^{20}\ \rm cm^{-2}$ in the source frame
(($2.70 \pm 0.20) \cdot 10^{20}\ \rm cm^{-2}$ for EPIC pn  and
 ($3.25 \pm 0.25) \cdot 10^{20}\ \rm cm^{-2}$ for EPIC MOS).
The soft X-ray absorption detected in MS 0737.9+7441 appears to be
fairly similar to other studies. Beckmann and Wolter (2000) found a mean
value for the intrinsic absorption at soft X-ray energies of about
$N_{\rm H, fit} = \rm 1.0 \cdot 10^{20}\ \rm cm^{2}$ with the largest value 
of $1.0 \cdot 10^{21}\ \rm cm^{-2}$.
The intrinsic absorption found in the host galaxy of the BL Lac MS 0737.9+7441 is also similar to the soft X-ray absorption of radio-quiet active galaxies
(cf. Table 1 of Boller, Brandt \& Fink 1996 and Table 2 of Walter \& Fink
1993). 
The flux variations of about 10 \% are relatively small for a BL Lac object 
(compare Giommi et al. 1990).
No significant spectral variations are detected during the observations. 

\acknowledgements
It is a pleasure to acknowledge the efforts of the SOC and SSC teams
in making the observations possible and for developing the SAS software
package used to reduce the data. We thank the referee, Eric Perlman, 
for very useful comments which helped us to improve the paper substantially.
The XMM - Newton project is supported by the Bundesministerium f\"ur
    Bildung und Forschung / Deutsches Zentrum f\"ur Luft- und Raumfahrt 
    (BMBF/DLR), the Max-Planck Society and the Heidenhain-Stiftung.
R.G.G and S.S. acknowledge the
support of PPARC, United Kingdom and S.V. acknowledges the support
of ASI, Italy.

\end{document}